\documentclass[journal=ancac3,manuscript=article,layout=twocolumn]{achemso}
\usepackage[version=3]{mhchem}
\usepackage{cleveref}
\usepackage{siunitx}
\usepackage{tabularx}

\title{Shaping Nanoscale Ribbons into Micro-Helices of Controllable Radius and Pitch }
\author{Lucas Prévost}
\affiliation[PMMH]
{Laboratoire de Physique et Mécanique des Milieux Hétérogènes (PMMH), UMR7636 CNRS, ESPCI Paris, PSL Research University, Sorbonne Université, Université de Paris, 75005 Paris, France}
\author{Dylan M. Barber}
\affiliation[Amherst]
{Polymer Science and Engineering Department, University of Massachusetts, Amherst, MA 01003, USA}
\author{Marine Daïeff}
\affiliation[PMMH]
{Laboratoire de Physique et Mécanique des Milieux Hétérogènes (PMMH), UMR7636 CNRS, ESPCI Paris, PSL Research University, Sorbonne Université, Université de Paris, 75005 Paris, France}
\author{Jonathan T. Pham}
\affiliation[Kentucky]
{Department of Chemical and Materials Engineering, University of Kentucky, Lexington, KY 40506, USA}
\author{Alfred J. Crosby}
\author{Todd Emrick}
\affiliation[Amherst]
{Polymer Science and Engineering Department, University of Massachusetts, Amherst, MA 01003, USA}
\author{Olivia du Roure}
\email{olivia.duroure@espci.fr}
\author{Anke Lindner}
\affiliation[PMMH]
{Laboratoire de Physique et Mécanique des Milieux Hétérogènes (PMMH), UMR7636 CNRS, ESPCI Paris, PSL Research University, Sorbonne Université, Université de Paris, 75005 Paris, France}


\setkeys{acs}{maxauthors=1, etalmode=truncate}
\begin{document}

% \begin{tocentry}
% \includegraphics{Figure_TOC.eps}
% \end{tocentry}

\begin{abstract}
We report fabrication of highly flexible micron-sized helices from nanometer-thick ribbons.
Building upon the helical coiling of such ultra-thin ribbons mediated by surface tension, we demonstrate that the enhanced creep properties of highly confined materials can be leveraged to shape helices into the desired geometry with full control of the final shape.
The helical radius, total length and pitch angle are all freely and independently tunable within a wide range: radius within $\sim \SIrange{1}{100}{\micro \meter}$, length within $\sim \SIrange{100}{3000}{\micro \meter}$, and pitch angle within $\sim \SIrange{0}{70}{\degree}$.
This fabrication method is validated for three different materials: poly(methyl methacrylate), poly(dimethylaminoethyl methacrylate), and transition metal chalcogenide quantum dots, each corresponding to a different solid-phase structure: respectively a polymer glass, a crosslinked hydrogel, and a nanoparticle array. 
This demonstrates excellent versatility with respect to material selection, enabling further control of the helix mechanical properties.
\end{abstract}

\section{Introduction}
Helical structures play a crucial role in fields spanning chemistry, biology, and mechanics. 
In particular, nature offers striking examples that span several orders of magnitude in length, including double-stranded DNA, $\alpha$-helix structures in proteins, cholesteric crystallization \cite{zastavker1999self}, or plant tendrils \cite{gerbode2012cucumber}.
Helices are of special importance for the motility of microorganisms, such as \textit{E. coli} or \textit{Salmonella} bacteria \cite{lauga2009hydrodynamics}, as many self-propel by rotating flexible helically-shaped flagella.
At sub-millimeter scales, viscous effects dominate and reciprocal motions cannot generate thrust \cite{purcell1977life} while helix rotation is non-reciprocal and allows propulsion. 
Similarly, synthetic materials and technologies rely on helical structures, including metallic nanosprings \cite{liu2014helical}, artificial micro-swimmers \cite{nelson2010microrobots}, or flow micro-sensors \cite{attia2009soft,li2012superelastic}.

Numerous methods have been reported for preparation of microscale helices, using either direct fabrication or spontaneous helix formation. 
Direct fabrication methods rely on complex manufacturing techniques, such as 3D-printing \cite{tottori2012magnetic,lu2018nanorobotic} or electro-spinning \cite{silva2017shaping}. For the spontaneous formation methods, a wide variety of mechanisms have been exploited: bilayer or trilayer systems \cite{li2012superelastic,zhang2009artificial,huang2005nanomechanical,
zhang2017dynamic,jeong2017topography,liu2014structural}, plastic deformation \cite{prior2016ribbon}, differential swelling \cite{douezan2011curling}, material anisotropy \cite{yu2017shape,zastavker1999self}, or surface tension \cite{pham2013highly}. 
    
While flexibility is often a key feature of natural helical structures, nearly all of these methods produce helices with limited flexibility.
To address this limitation, spontaneous formation methods have been discovered and developed to provide flexible structures \cite{zastavker1999self,li2012superelastic,pham2013highly,zhang2017dynamic,maier2017self}; however, they tend to  lack precise control of the helix shape and dimensions.
Specifically, the helical radius and pitch frequently cannot be freely and independently tuned.
Li \textit{et al.} \cite{li2012superelastic} did report fabrication of flexible metallic microsprings with full geometric control, though the material choice was restricted to metal composites so the resulting helices are fairly rigid, requiring significant stress to observe deformation.
Maier \textit{et al.} \cite{maier2017self} reported self-assembly of helical DNA nanotubes with full geometric control, but the material choice is restricted to DNA and the flexibility of these structures has yet to be characterized.
Overall there remains a crucial need for a versatile fabrication method of flexible micro-helices of tunable geometry and well-characterized flexibility. 

Here, we address this challenge by implementing a two-step fabrication method for flexible helices made from nanometer-thick ribbons.
Our method builds upon a spontaneous formation technique driven by surface tension \cite{pham2013highly} and harnesses the creep properties of materials to shape helices into the desired geometry.
We demonstrate direct and local control of the helical pitch.
As control of the helical radius and total filament length was already achieved, the helix geometry is fully tunable.
This fabrication method is validated for three different materials: poly(methyl methacrylate), poly(dimethylaminoethyl methacrylate), and semi-conductor quantum dots. 
Each of these three materials exhibits a different solid-phase structure, respectively a polymer glass, a crosslinked hydrogel, and a nanoparticle array, demonstrating versatility in the material choice. 
The resulting helices are micron-sized in radius, slender, and highly flexible. 

\section{Results and Discussion}

\subsection{Principles of Fabrication}
\begin{figure*}[t!]
  \centering \includegraphics[width= 1\textwidth]{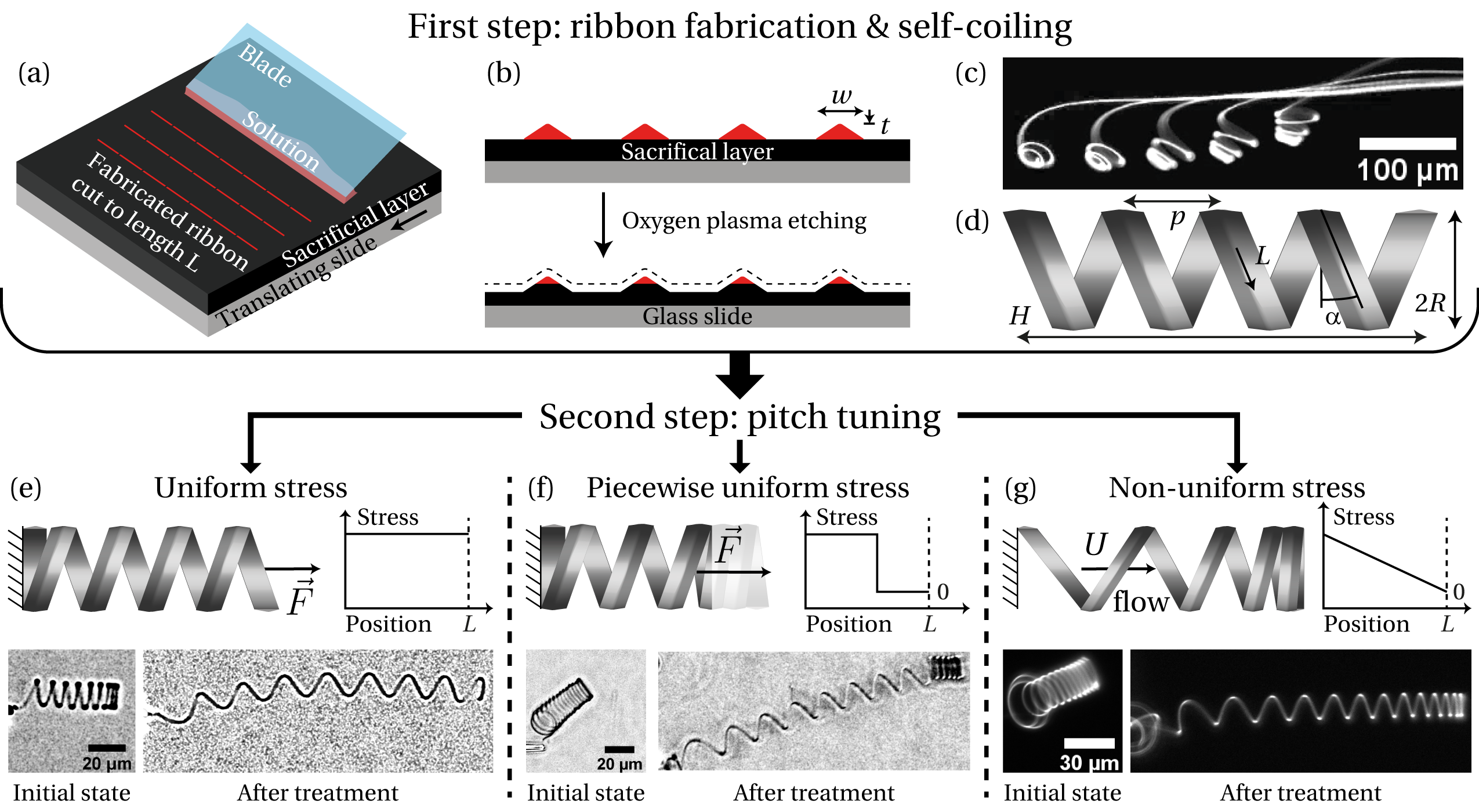}
  \caption{Fabrication work-flow: firstly ribbons are fabricated and coiled into helices; secondly, the pitch is encoded. (a) Flow-coating method for ribbon fabrication. Fabrication settings control the obtained thickness and width. The total length $L$ can be tuned during the fabrication or by cutting. (b) Oxygen plasma etching of the fabricated ribbons. The thickness and width can be further tuned, typical values are $w \sim \SIrange{0.5}{5}{\micro \meter}$ and $t \sim \SIrange{10}{50}{\nano \meter}$. (c) Chrono-photography of a helical ribbon forming in water. The time step between pictures is $\SI{2}{\second}$. (d) Helical ribbon geometry and relevant parameters. (e) to (g): different methods for pitch tuning, depending on the stress applied, either (e) uniform stress, (f) piecewise uniform stress or (g) non-uniform stress. For each method we show a schematic of the method, a representation of the stress along the filament, and the corresponding before/after experimental images. Images are taken using bright-field microscopy (light background) or fluorescent microscopy (dark background).
  }
  \label{fig1}
\end{figure*}

The first step of our two-step method relies on the spontaneous formation of highly-flexible helical ribbons driven by surface tension \cite{pham2013highly}. 
Nanometer-thick ribbons are prepared on a flat sacrificial layer through an evaporative assembly method termed flow-coating \cite{kim2010nanoparticle} (\cref{fig1}.a). 
As a result of the fabrication process, the ribbons display a near-triangular cross-section of width $w$ (typically $\SIrange{0.5}{5}{\micro \meter}$) and thickness $t$ (typically $\SIrange{10}{50}{\nano \meter}$) such that $t \ll  w$, as depicted in \cref{fig1}.b.
Upon release of the ribbons into a liquid, they spontaneously coil into a helical geometry. 
We show in \cref{fig1}.c a chrono-photography of the coiling process for a helical ribbon in water. 

The driving force of the coiling is the reduction of surface area, and hence of surface energy, induced by bending of the ribbon. 
This is a consequence of the ribbons' cross-sectional asymmetry. 
There is, however, no energetic incentive to twisting the filament, favoring closed-loop helices.
The resulting helical ribbon is described by its total length $L$, helical pitch $p$ and helical radius $R$ (see \cref{fig1}.d). 
The helix pitch angle $\alpha$ is defined such that $\tan \alpha = p/2\pi R$ and the helix axial length $H$ is defined such that $H = L \sin \alpha$.

The filament length $L$ is directly controlled during the fabrication process and can be further tuned by cutting the ribbons prior to release in liquid. 
The helical radius $R$ is determined by a balance between surface tension and elasticity and scales as $R \sim Et^2/\gamma$, with $E$ being the Young's modulus and $\gamma$ the surface tension \cite{pham2013highly}. 
This affords control over the radius through modification of the ribbon thickness, either during ribbon fabrication or by their further etching \cite{choudhary2019controlled} (see \cref{fig1}.b). 
Furthermore, the radius can be varied over a very wide range, within $\sim \SIrange{1}{100}{\micro \meter}$.
As the preferred filament torsion is zero, the helical ribbons typically adopt tight configurations. 
The reference pitch of the helix (\textit{i.e.} when no external stress is applied) is thus determined by contact between the successive loops and is then of the order of the ribbon width $w$.
Control of the helical pitch is therefore lacking but the processes introduced below overcome this limitation.

We observed that when stress is exerted on a helical ribbon, some deformation persists in the structure even after the stress is relaxed.
These irreversible deformations are observed for all tested materials and enable in-situ tuning of the helical pitch, after release and coiling of the helices.
The process creates a persistent stress in the material by extending the helix for a long period of time, typically several minutes, after which the helix is allowed to relax. 
Different stress profiles can be applied  to locally control the final helix dimensions, as shown in \cref{fig1}.e to \cref{fig1}.g: uniform stress, piecewise uniform stress or non-uniform stress.

Uniform stress along the ribbon length is achieved by loading the whole helix (\cref{fig1}.e).
Piecewise uniform stress is achieved by loading only a fraction of the helix (\cref{fig1}.f).
Finally a gradient stress profile is realized by immersing the helix in an axial viscous flow (\cref{fig1}.g).
For each stress profile, a different irreversible deformation is obtained, which is discussed in more detail below. 
These pitch control methods cannot be used to decrease the pitch since inward forces buckle the helix instead of compressing it. 
But as helices intrinsically display vanishing pitch, arbitrary values for the helical pitch can be reached nevertheless. 

\subsection{Experimental Set-Up}

The ribbons are prepared on a water-soluble sacrificial layer, following the method established by Lee \textit{et al.} \cite{lee2013macroscopic}, then released into a pool of water connected to a microfluidic channel (see \cref{fig2}.a for the full experimental set-up) prepared from polydimethylsiloxane (PDMS) using standard soft lithography techniques. 
Two open glass capillary tubes are connected to syringe pumps and fixed to two micromanipulators to allow capture and release of the helices by pumping or expelling liquid.

Our process was effective across diverse ribbon compositions, enabling further control of the helix mechanical properties.
We tested poly(methyl methacrylate) (PMMA), for which most of the results are presented.
The PMMA ribbons are prepared on a sacrificial layer of poly(acrylic acid) (PAA) and are not crosslinked. 
We also tested poly(dimethylaminoethyl methacrylate) (PDMAEMA) prepared on a sacrificial layer of poly(styrene sulfonate) (PSS). 
The PDMAEMA ribbons are crosslinked. 
Finally, we tested CdSe quantum dots (QDs) prepared on a PAA sacrificial layer.

Upon release into water, the ribbons quickly adopt (within $\sim$ one minute) a tight helical configuration. 
One end of the helix is caught and clamped by the first capillary, termed the 'holder', by pumping liquid into the capillary. 
We can then either approach the second capillary, termed the 'puller', in order to exert loading, or position the helix  inside the microfluidic channel, in order to exert viscous forces.
All experiments are conducted at room temperature $\sim \SI{22}{\celsius}$.

\subsection{Mechanics of Helical Ribbons}

\begin{figure}[b!]
  \centering \includegraphics[width= 1\linewidth]{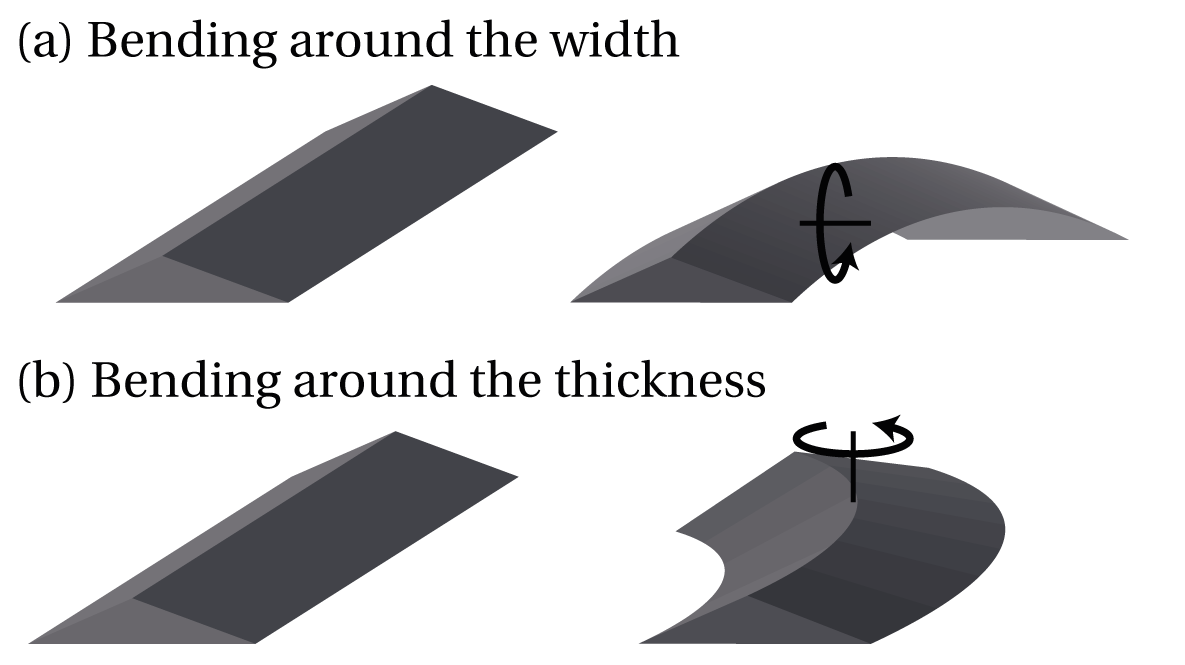}
  \caption{Schematic of the two modes of ribbon bending. (a) Bending around the width, the corresponding bending modulus is $B$. (b) Bending around the thickness.}
  \label{fig6}
\end{figure}

Before detailing the different processes, we introduce a basic mechanical framework describing the mechanics of helical ribbons.
The helical structure is highly slender ($R/t \sim 300-1000$) so its deformation is dominated by bending and twisting of the ribbon. 
The corresponding filament moduli are $B$ bending modulus and $C$ twisting modulus. 
$B$ quantifies the resistance of the ribbon to bending around its width (as illustrated in \cref{fig6}.a) while $C$ quantifies the resistance of the ribbon to twisting, \textit{i.e.} rotation around its length. 
Both scale similarly $C \sim B \sim E w t^3$ but are not equal. 
More precisely, for a flat triangular cross-section, we have $B = \displaystyle \frac{1}{36} E w t^3$ and $C = \displaystyle \frac{1}{12} \mu w t^3$ with $\mu$ shear modulus \cite{gloumakoff1964}.

Because the ribbon is flat, no bending occurs around the thickness direction (deformation mode illustrated in \cref{fig6}.b).
In this case, the material frame matches the Frenet frame of the centerline \cite{mahadevan1993shape,starostin2007shape}. 
The curvature and torsion of the material are hence easily computed as the Frenet curvature and torsion of the centerline.
And thus, the local bending stress $\sigma_\text{bend}$ (with $B$ the corresponding modulus) and local torsional shear $\tau_\text{shear}$ (with $C$ the corresponding modulus) can be expressed from the geometry of the centerline. 
Their maximum values on the cross-section are respectively: $\sigma_\text{bend} = \frac{2}{3} E t \Delta \kappa$ and $\tau_\text{shear} = \mu t \Delta \tau$ \cite{gloumakoff1964}.
$\Delta \kappa$ is the local change (between the deformed state and the reference state) of Frenet curvature while $\Delta \tau$ is the local change of Frenet torsion.
For a uniform helix (uniform radius and angle), the centerline curvature is $\kappa = \cos^2\alpha/R$ and the centerline torsion is $\displaystyle \tau = \cos \alpha \sin \alpha /R$.
Hence changes in Frenet curvature or torsion scale similarly $\Delta \kappa \sim \Delta \tau \sim 1/R$.
And thus, the bending stress and torsional shear also scales similarly $\sigma_\text{bend} \sim \tau_\text{shear} \sim E \ (t/R)$.

Specific to the case of the end-loading (depicted on \cref{fig1}.e), we introduce the axial force $F$, meaning the force necessary to stretch the helix. 
The tension force scales as $F \sim (C/R^2) \times (\Delta H/L)$ with $\Delta H$ the change in axial length \cite{love2013treatise}.
The tensile stress $\sigma_\text{tens}$ scales as $\sigma_\text{tens} \sim F/wt \sim E \ (t/R)^2$.
Given that $t \ll R$, we have $\sigma_\text{tens} \ll \sigma_\text{bend}, \tau_\text{shear}$: bending and twisting effects are indeed dominant. 
The exact value of the tension force depends in a complex way on the helical geometry and boundary conditions. The work of Starostin \textit{et al.} \cite{starostin2008tension} provides an analytical expression 
\begin{equation}
F = \frac{C}{R^2} \frac{\Delta H}{L} \frac{1}{\cos^2 \alpha \left( \cos^2 \alpha + (C/B) \sin^2 \alpha \right) }
\label{eq1}
\end{equation}
We now detail the three different processes, each associated to a different stress profile imposed during the treatment: uniform stress, piecewise uniform stress, or non-uniform stress.

\subsection{Uniform Stress: the Stretching Treatment}

\begin{figure*}[t!]
  \centering \includegraphics[width= 1\linewidth]{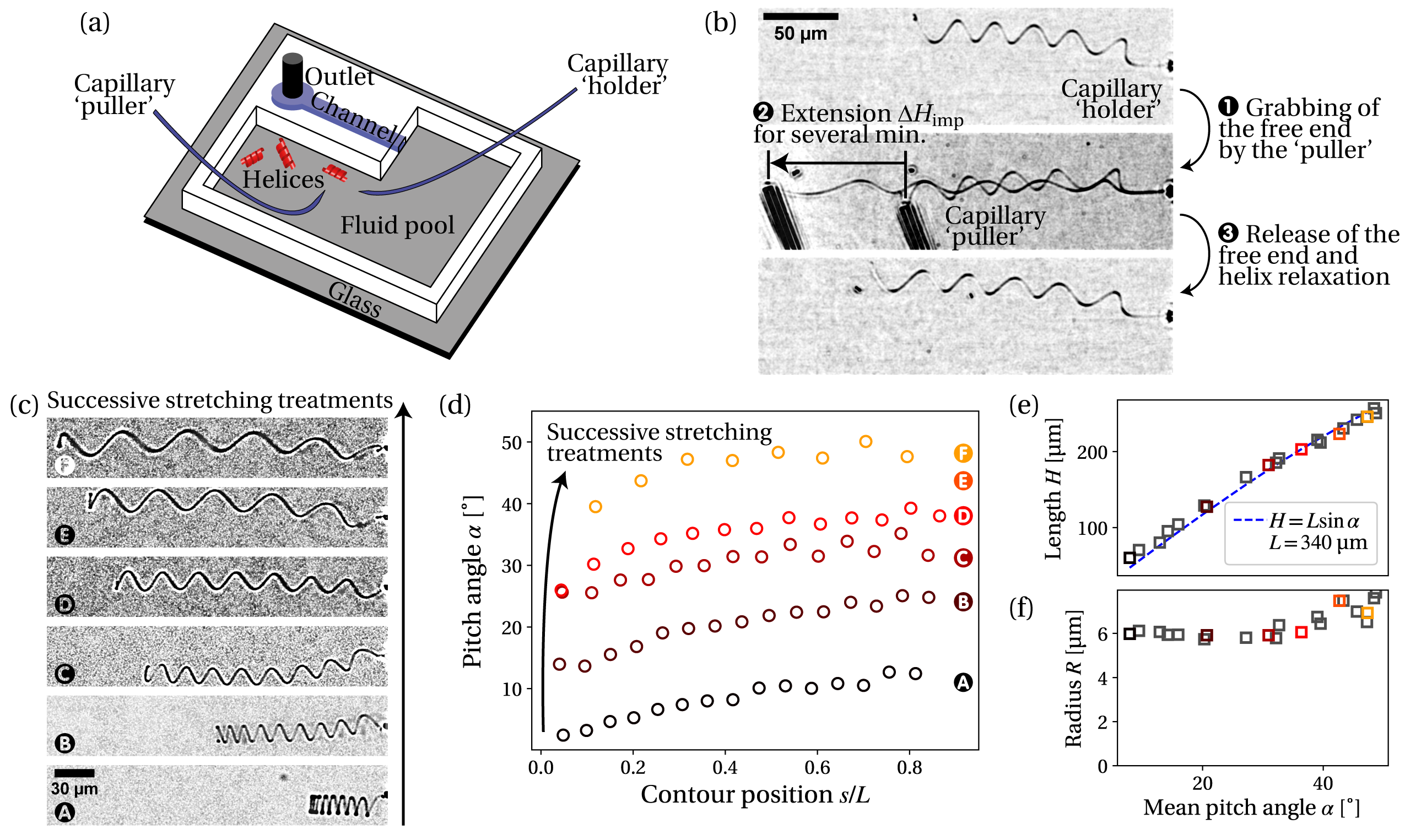}
  \caption{(a) Schematic of the experimental set-up. (b) Images illustrating the uniform pitch increase process through end-loading, termed stretching treatment. The helix right end is clamped by the holder and the left end is grabbed by pumping liquid into the puller. The helix is then extended by displacing the puller capillary. After several minutes, the left end is released by expelling liquid out of the puller capillary. (c) One PMMA helix after several successive stretching treatments and (d) corresponding pitch distribution along the filament length. (e) Axial length as a function of mean helix pitch angle following the successive stretching treatments, with fitting. (f) Evolution of the mean helical radius as a function of the mean helix pitch angle.}
  \label{fig2}
\end{figure*}

The uniform stress method, termed 'stretching treatment', is done in three steps depicted in \cref{fig2}.b. 
First, while one end of the helical ribbon is held in place by the holder capillary, the free end is grabbed by the puller capillary.
We then impose to the helix a fixed axial extension $\Delta H_\text{imp}$ for several minutes by displacing the puller.
Finally, the free end is released from the puller by expelling liquid out of the capillary. 
This results in a permanent increase in the helical pitch and thus in the axial length (compare the top and bottom images of \cref{fig2}.b). 
The resulting change in axial length is noted $\Delta H_\text{res}$. 
To compare different helices more easily, we track changes in pitch angle $\alpha$, instead of pitch, as $\alpha$ is expressed as a function of the rescaled pitch, $p/R$.

We show in \cref{fig2}.c the changes resulting from successive stretching treatments applied to a PMMA helix.
The mean pitch angle gradually increases from 8° to 48°. 
Around twenty stretching treatments were performed during this experiment, with a selection of experimental images shown in \cref{fig2}.c.
Three stretching treatments were performed  between each image.
The corresponding angle distributions along the filament length are shown in \cref{fig2}.d. 
The increase in pitch angle is uniform along the length and preexisting heterogeneities tend to be relaxed. 
This is consistent with the fact that, for end-loading, the elastic stress and the elastic deformation are uniform along the filament length.

For this same experiment, the evolution of the axial length $H$ as a function of the mean pitch angle is plotted in \cref{fig2}.e, showing the full evolution of the helix, after each stretching treatment.
The data points corresponding to the images shown in \cref{fig2}.c are highlighted.
The evolution of $H$ is accurately fitted by the expected geometrical relation $H = L \sin \alpha$, meaning that the total filament length $L$ is kept constant throughout the experiment.
We note that the helical radius evolves slightly as a side effect of the treatment. 
The evolution of the radius as a function of the pitch angle is plotted in \cref{fig2}.f.
The radius change is negligible until $\sim \SI{30}{\degree}$ pitch angle, and then increases up to $\sim 25 \%$ of its initial value. 
This change in helical radius causes a change in the number of turns, as seen in \cref{fig2}.c.

\begin{figure*}[t!]
  \centering \includegraphics[width= 1\linewidth]{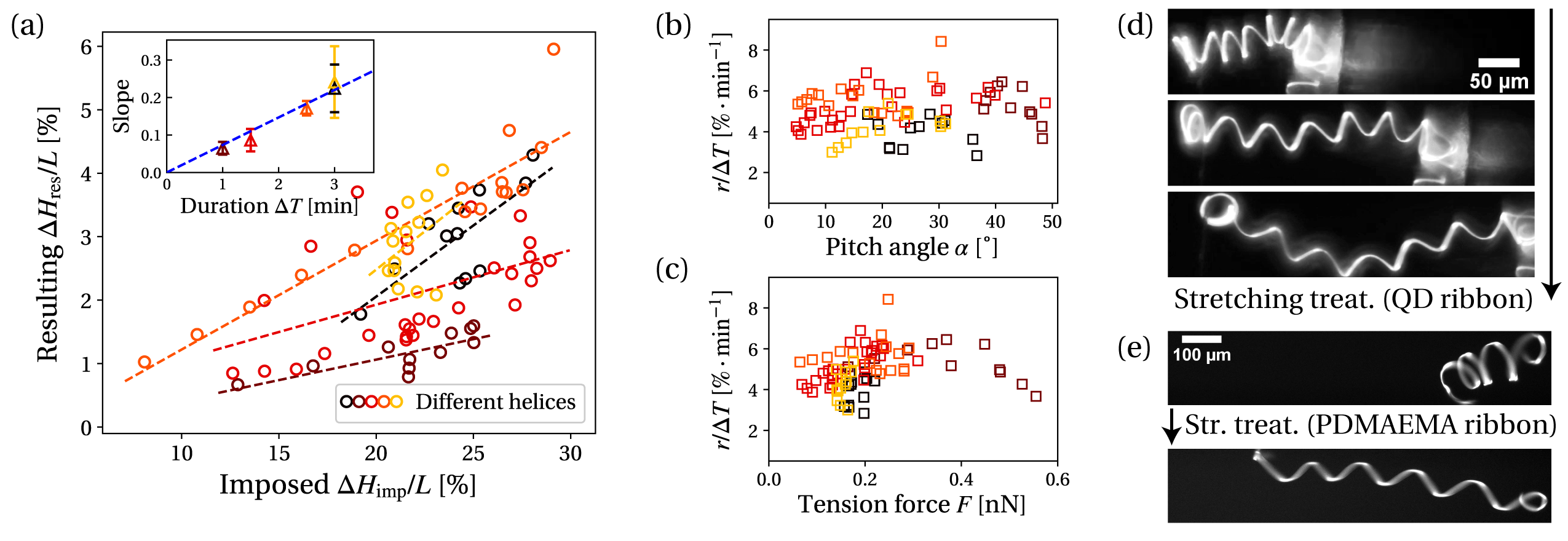}
  \caption{Results of successive stretching treatments applied to several PMMA helices: (a) resulting increase in axial length $\Delta H_\text{res}$ as a function of imposed axial stretching $\Delta H_\text{imp}$, both rescaled by the total length $L$. 
  Dashed colored lines are linear fitting. 
  Inset shows the slope as a function of the treatment duration $\Delta T$, with linear fitting (blue dashed line). 
  (b) Time-corrected ratio $r/\Delta T$ (with $r = \Delta H_\text{res} / \Delta H_\text{imp}$) as a function of helix pitch angle and (c) as a function of the force $F$ applied to stretch the helix.
  (d) QD helix in its initial state and after two successive stretching treatments. 
  (e) PDMAEMA helix before and after a stretching treatment.}
  \label{fig4}
\end{figure*}

To describe more accurately the stretching treatment, we measure the resulting increase in axial length $\Delta H_\text{res}$ as a function of the imposed axial elongation $\Delta H_\text{imp}$ for several PMMA helices. 
The duration of the treatment $\Delta T$ is kept constant for a given helix but varies from helix to helix. 
For all tested helices, the resulting increase in axial length is linear with the imposed stretching, as shown in \cref{fig4}.a.
Furthermore, the slope is linear with the treatment duration $\Delta T$ (see inset in \cref{fig4}.a). 
The resulting increase in axial length is thus both linear with time and with the imposed stretching.
Naturally, we expect the axial length increase to saturate at very long stretch time, but this saturation effect was not observed within our range of treatment duration, \textit{i.e.} up to $\SI{3}{\minute}$.
The ratio $r = \Delta H_\text{res} / \Delta H_\text{imp}$ divided by the duration time $\Delta T$ is therefore roughly constant throughout all PMMA helices: $r/\Delta T = \SI{5.0 \pm 1.0}{\percent \per \minute}$. 

Notably, this time-corrected ratio does not vary with pitch angle, as shown in \cref{fig4}.b, using the pitch angle as a proxy for time and number of applied treatments. 
The angle can indeed only increase during the experiment, and the increase is a consequence of the applied stretching treatments. 
The non-correlation with the pitch angle means that neither time nor repeated treatments influence the stretching treatment.
We also check that the time-corrected ratio does not vary with the tension force applied, as shown in \cref{fig4}.c. 
The tension force is calculated using the analytical expression proposed by Starostin \textit{et al.} \cite{starostin2008tension}.
The time-corrected ratio is therefore characteristic of the material, independent of the helix geometrical parameters.

Finally, we calculate the typical stresses (bending stress and torsional shear) imposed during treatment.
As mentioned, these stresses can be computed from the centerline geometry, which is easily obtained from the experimental images. 
For both stresses, we obtain vanishing values: we have $\sigma_\text{bend}/E = \SIrange{d-4}{d-3}{}$ and similarly $\tau_\text{shear}/\mu = \SIrange{d-4}{d-3}{}$. 

The stretching treatment, yielding uniform pitch increase, gives similar results for the other tested materials: PDMAEMA and quantum dots. 
\Cref{fig4}.d shows an example of two successive stretching treatments applied to a QD helix. 
We consistently obtain a uniform increase in the helical pitch. 
For QD helices, we estimate the time-corrected ratio to $r/\Delta T \sim \SI{7}{\percent \per \second}$ (notice the change in the time unit). 
The value of the time-corrected ratio is significantly higher than for PMMA helices.
Similarly, we show in \cref{fig4}.e the result of one stretching treatment applied to a PDMAEMA helix, estimating the time-corrected ratio at an intermediate value $r/\Delta T \sim \SI{1}{\percent \per \second}$.

As evidenced by the PMMA helical ribbons, control of the stretching treatment is very convenient: the increase in axial length is directly proportional to the imposed stretching and to the treatment duration. 
The effectiveness of the treatment is not affected by time nor by repetition.
Also, conveniently, the stretching treatment can be performed in minutes. 
The resulting helices retain their characteristic flexibility, comparable to the flexibility of bacterial flagella \cite{pham2015deformation}.
We can thus fabricate well-controlled highly flexible helical ribbons of arbitrary pitch and radius. 
These helical ribbons may act as model systems for natural helical structures such as bacterial flagella, with added control on the geometrical and mechanical parameters, beyond biological limitations.
Detailed experimental investigations of the behavior of flexible helices can thus be conducted in various situations, such as polymorphic transformations \cite{darnton2007force} or free transport in flow \cite{li2021tumbling}.

The value of the time-corrected ratio is important in regards to these possible applications. 
On the one hand, a vanishing value would make the stretching treatment very slow and would thus render the fabrication process inconvenient.
One the other hand, a high value would make the helical ribbons unusable as a model system:  any stress applied to the helix would deform it irreversibly within seconds.
A low value of a few percent per minute is desirable. 
The stretching treatment can be performed in reasonable time (several minutes), and irreversible deformations can be neglected during other experiments, as long as experiment duration does not exceed a few minutes.
PMMA helices display the most favorable time-corrected ratio for these applications.

The limit on the possible experiment duration is the main drawback of this fabrication method. 
A controllable and in-situ technique to slow or stop these irreversible deformations would greatly benefit the method. 
Cross-linking the material does not seem to prevent these effects, as shown by the case of PDMAEMA helices, although the case of cross-linked PMMA was not studied. 
Cooling the system may slow the deformation dynamic. 

\subsection{Towards More Complex Geometries}

\begin{figure*}[t!]
  \centering \includegraphics[width= 1\linewidth]{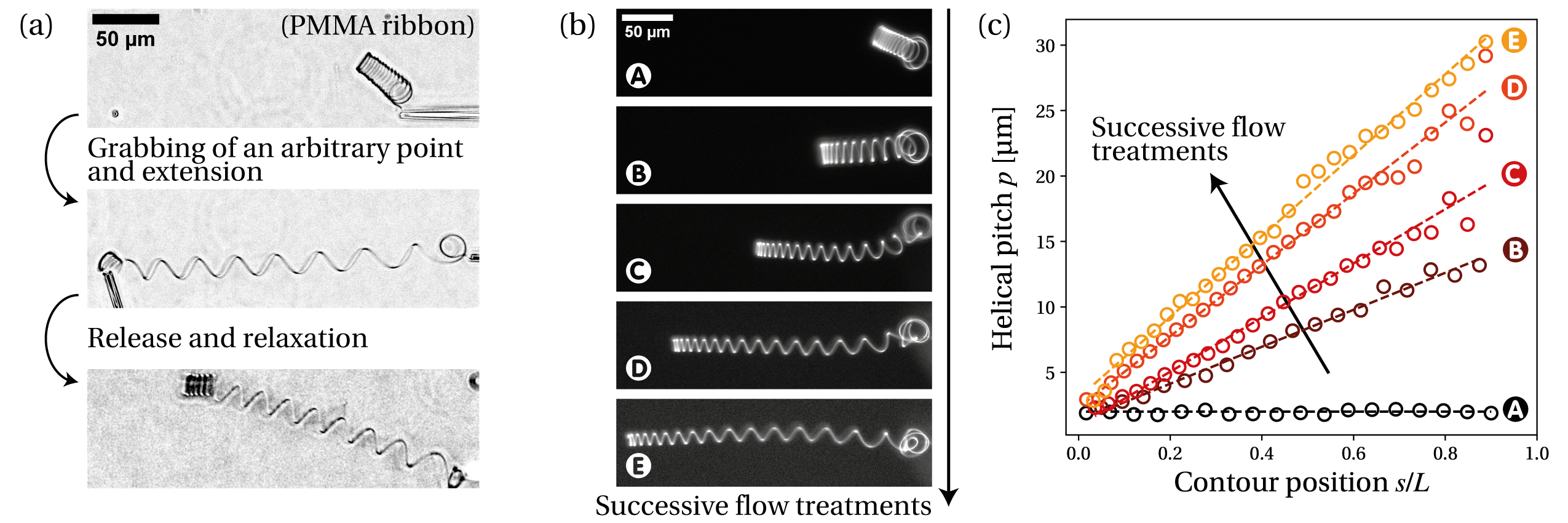}
  \caption{(a) Fabrication of a two-part PMMA helix by selectively applying the stretching treatment, here only on the right part of the helix. (b) PMMA helix after several successive flow treatments and (c) corresponding pitch distribution along the filament length, with linear fit (dashed lines). $s=0$ corresponds to the free end while $s=L$ corresponds to the clamped end.}
  \label{fig5}
\end{figure*} 

As shown in the previous section, end-loading the helix increases the pitch uniformly along the whole filament. 
Consistent with these observations, loading only a fraction of the helix yields a uniform pitch increase only in the loaded part.
We illustrate this in \cref{fig5}.a using a PMMA helix, in which only a portion of the helix is stretched (roughly the right half).
The stress applied is then piecewise uniform. 
This results in a two-part helix, each part displaying uniform pitch distribution but with different values. 

This two-part helix may be viewed as a rough model of a uni-flagellated bacteria: a chiral helically shaped flexible flagella and a non-chiral cylindrical body, with good control of the geometrical parameters.
The two-part helix may also be used as a force sensor with extended measurement range.
The pitch angle has a strong influence on the spring constant of a helical ribbon (see \cref{eq1}): increasing the pitch angle effectively stiffens the helix. 
The two-part helix is hence composed of a softer sensor (small angle part) and a stiffer sensor (high angle part), effectively increasing the measurement range. 

The helix can finally be loaded by a non-uniform stress, achieved by immersing the helix in a uniform viscous flow.
For this method, termed 'flow treatment', the helix is positioned at the center of the microfluidic channel and a flow is applied from the clamped end towards the free end, extending the helix. 
As the helix size is small compared to the channel size, this is equivalent to immersing the helix in a uniform flow. 
After several minutes, the flow is stopped and the helix is allowed to relax. 

\Cref{fig5}.b shows the result of several successive flow treatments applied to a PMMA helix. 
The corresponding pitch distributions, plotted in \cref{fig5}.c, are accurately fitted by a linear function along the filament length.
The pitch increase is maximum at the clamped end ($s/L =1$) and zero at the free end ($s/L = 0$). 
The pitch increase distribution matches the distribution of the elastic deformation imposed to the helix during the flow treatment, which can be approximated to a linear function of the contour position \cite{pham2015deformation}. 
This confirms that, similarly to the stretching treatment, the resulting deformation is directly proportional to the imposed elastic stretching.

Control of the flow treatment is more complicated than for the stretching treatment. 
The deformation imposed during flow is force-controlled rather than displacement-controlled, since we impose the velocity of the flow and thus the viscous forces acting on the helix. 
Also, precise calculation of the viscous forces acting on the helix is not straightforward, requiring the use of numerical tools such as slender-body theory \cite{rodenborn2013propulsion}.

As evident from our observations, these helical ribbons sense the flow history imposed upon them.
The total length increase tracks the total flow volume, and the pitch increase distribution gives information on the local flow geometry. 
For example, a uniform flow results in a linear increase of the helical pitch.
For non-uniform flows, the pitch increase distribution will simply match the distribution of the elastic stretching imposed by the flow.
The different materials can be used to tune the sensor sensitivity: QD helices for short times or weak flows or PMMA helices for long times or strong flows. 

\subsection{Discussion}
The irreversible deformations that we observe happen under persistent stress at slow rate.
They occur at very low stress, significantly below the material yield stress: the typical values obtained experimentally for the local stresses are at most $\tau_\text{shear}/\mu \sim 10^{-3}$. 
These small local stress values are obtained despite significant global displacement, a consequence of the slender nature of the filament. 
The process is linear and unaffected by time or repeated treatments. 
It is therefore likely that material creep is the physical phenomenon at play, hence the mechanical properties will be similar to those for other polymer glasses in terms of frequency, time, and temperature responses \cite{ferry1980viscoelastic}.

Considering the low temperature ($\sim \SI{22}{\celsius}$) and vanishing stress values, the sizable creep effects we observe are not expected for bulk materials but may develop in highly confined nanoscale geometries. 
In the case of highly confined glassy polymers, such as PMMA, recent studies have indeed highlighted a sharp decrease in the glass transition temperature \cite{roth2005glass,pye2011two,o2008creep,mattsson2000quantifying} as well as a overall decrease in viscosity \cite{rowland2008measuring,bodiguel2006reduced,li2013viscosity}.
These observations are usually interpreted as a consequence of a loss of interchain entanglements \cite{sussman2014entanglement,si2005chain} and an increased mobility of the polymer chains near the free surface \cite{martin2010direct}, which might explain the enhanced creep behavior of PMMA that is taken advantage of here.
A more detailed study of this behavior would be beneficial to the understanding of the rheology of highly confined materials. 
This would be also valuable in the case of QD and PDMAEMA, materials for which analysis of the creep properties is lacking in the literature and for which a detailed characterization was not performed in this work.

\section{Conclusions}
This work has presented a fabrication method for micron-sized highly flexible helical ribbons, with full control on the helix dimensions.
Helix formation and control of the helical radius and total length rely on a previously established method \cite{pham2013highly}. 
We have added pitch control by taking advantage of the creep properties of materials.
Sizable creep deformations are not expected in bulk but are enabled by the nanoscale thickness of the ribbons.

The in-situ processes presented allow for a direct, local and precise tuning of the helical pitch.
The so-called stretching treatment creates a uniformly distributed helical pitch, by end-loading the helix. 
Similarly, localized loading of the helix yields a uniform increase in pitch only in the loaded part. 
The so-called flow treatment creates a linearly distributed helical pitch, by immersing the helix in a uniform viscous flow. 
In all cases, the resulting pitch is directly proportional to the imposed deformation and to the treatment duration, with no influence of time or repeated treatments.
Control of the pitch is therefore straightforward. 
All other helix properties (geometrical or mechanical) remain unaffected, with the exception of the helical radius, which can slightly vary from its initial value during treatment.

This fabrication method was validated for three different materials: poly(methyl methacrylate), poly(dimethylaminoethyl methacrylate) and quantum dots, each corresponding to a different solid-phase structure ; respectively a polymer glass, a crosslinked hydrogel and a nanoparticle array. 
This demonstrates versatility in the material choice, enabling further control of the helix mechanical properties.
Each material was characterized by a time-corrected ratio, which quantifies its susceptibility to creep. 
PMMA helices were found to be the least susceptible to creep, while QD helices were found to be the most susceptible to creep. 
PDMAEMA exhibited an intermediate susceptibility.

These helical ribbons can serve as model systems of natural helical structures, such as flagellated bacteria.
Coupled with microfluidic flow control techniques or cantilever force measurements, they allow the study of the complex physics of helical structures.
Finally they may benefit technological applications, such as force sensors with extended measurement range, history-sensitive flow sensors, or deformable micro-swimmers.

\section{Material and Methods}
\textit{Material and sample preparation}
Glass slides (of size $\SI{24}{\milli \meter} \times \SI{40}{\milli \meter} \times \SI{170}{\micro \meter}$) are sonicated for 15 minutes successively in soap + water, water and isopropanol and dried with compressed air. 
The slides are treated with oxygen plasma (Plasma Diener Pico) for 2 minutes before spin-coating (spin-coater POLOS Spin150i) $\SI{25}{\milli \gram \per \milli \liter}$ polyacrylic acid (PAA, $M_\text{w} = 1800$, Sigma-Aldrich) aqueous solution at 3000 rpm (rotations per minute) for $\SI{30}{\second}$.
Poly(methyl methacrylate) (PMMA, $M_\text{w} = \SI{120d3}{}$, Sigma-Aldrich) is suspended in anhydrous toluene ($\SI{99.8}{\percent}$, Sigma-Aldrich) at $\SI{1}{\milli \gram \per \milli \liter}$ concentration with minimum amount of Coumarin 153 (Sigma-Aldrich) added to afford fluorescence. 
Details on the preparation of the poly(dimethylaminoethyl methacrylate) (PDMAEMA) samples can be found in a previous publication \cite{barber2020programmed}.
Details on the preparation of the CdSe quantum dot (QD) samples can be found in a previous publication \cite{pham2013highly}.

\textit{Flow coating}
The flow-coating setup, described in detail in the literature \cite{kim2010nanoparticle,pham2013highly}, consists of a translation stage (Newport LTA actuator), on which the coated glass slides are fixed. 
An angled silicon blade attached to a stage with pitch and roll variability (Newport M-37) is positioned above the slide.
$\SIrange{3}{5}{\micro \liter}$ of solution is loaded in the wedge between the blade and the substrate. 
The translation stage performs at stop-and-go motion, with stage velocity $\SI{5}{\milli \meter \per \second}$, stopping time $\SIrange{1.0}{4.0}{\second}$ and spacing $\SI{1.0}{\milli \meter}$. 
Flow-coated samples are etched using oyxgen plasma treatment (Plasma Diener Pico) to adjust the ribbon thickness. 
Ribbon thickness is measured by optical profilometry (Veeko Instruments Wyko NT9100).

\textit{Capillary fabrication}
The glass capillaries are prepared from standard glass tubes with $\SI{1}{\milli \meter}$ outside diameter and $\SI{0.58}{\milli \meter}$ inner diameter, using a micropipette puller system (P-1000 Flaming/Brown, Sutter). 
This creates a very thin closed tip of typical diameter $\sim \SI{1.5}{\micro \meter}$.
The closed tip is then melted to open the capillary, using a heated glass bead (MF-830 Microforge, Narishige International).
The final tip diameter of the open glass capillary ranges within $\SIrange{5}{20}{\micro \meter}$.
The tip of each capillary is coated with bovine serum albumin (BSA) by immersion in a $\SI{2}{\percent}$ BSA aqueous solution for 15 minutes.

\textit{Experimental set-up}
The experimental set-up is mounted on an inverted optical microscope (Zeiss Axio Observer) connected to a numerical camera (Hamamatsu Orcaflash LT 4.0). 
A UV-source is used for fluorescence (HXP 120 lamp, Zeiss) combined with a filter set matching the Coumarin 153 excitation bandwidth (filter set 40, Zeiss).
UV light influx is controlled by a shutter (Shutter Uniblitz V25).
The flow rate in the microchannel is controlled by a syringe pump (NeMESYS, Cetoni). 
The open glass capillaries are mounted on two micromanipulators (TransferMan, Eppendorf). 

\section{Acknowledgments}
LP and AL acknowledge funding by the European Research Council through a consolidator grant (ERC PaDyFlow 682367). 
AJC acknowledges support by, or in part by, the U. S. Army Research Laboratory and the U. S. Army Research Office under contract/grant number W911NF-15-1-0358.
DMB acknowledges a National Defense Science and Engineering Graduate Fellowship.
This work received the support of Institut Pierre-Gilles de Gennes (Équipement d’Excellence, “Investissements d’Avenir”, Program ANR-10-EQPX-34).

\section{Conflict of Interest}
The authors declare no conflict of interest.

\bibliography{bib_fabrication}

\providecommand{\latin}[1]{#1}
\makeatletter
\providecommand{\doi}
  {\begingroup\let\do\@makeother\dospecials
  \catcode`\{=1 \catcode`\}=2 \doi@aux}
\providecommand{\doi@aux}[1]{\endgroup\texttt{#1}}
\makeatother
\providecommand*\mcitethebibliography{\thebibliography}
\csname @ifundefined\endcsname{endmcitethebibliography}
  {\let\endmcitethebibliography\endthebibliography}{}
\begin{mcitethebibliography}{46}
\providecommand*\natexlab[1]{#1}
\providecommand*\mciteSetBstSublistMode[1]{}
\providecommand*\mciteSetBstMaxWidthForm[2]{}
\providecommand*\mciteBstWouldAddEndPuncttrue
  {\def\EndOfBibitem{\unskip.}}
\providecommand*\mciteBstWouldAddEndPunctfalse
  {\let\EndOfBibitem\relax}
\providecommand*\mciteSetBstMidEndSepPunct[3]{}
\providecommand*\mciteSetBstSublistLabelBeginEnd[3]{}
\providecommand*\EndOfBibitem{}
\mciteSetBstSublistMode{f}
\mciteSetBstMaxWidthForm{subitem}{(\alph{mcitesubitemcount})}
\mciteSetBstSublistLabelBeginEnd
  {\mcitemaxwidthsubitemform\space}
  {\relax}
  {\relax}

\bibitem[Zastavker \latin{et~al.}(1999)Zastavker, Asherie, Lomakin, Pande,
  Donovan, Schnur, and Benedek]{zastavker1999self}
Zastavker,~Y.~V. \latin{et~al.}  Self-assembly of helical ribbons.
  \emph{Proceedings of the National Academy of Sciences} \textbf{1999},
  \emph{96}, 7883--7887\relax
\mciteBstWouldAddEndPuncttrue
\mciteSetBstMidEndSepPunct{\mcitedefaultmidpunct}
{\mcitedefaultendpunct}{\mcitedefaultseppunct}\relax
\EndOfBibitem
\bibitem[Gerbode \latin{et~al.}(2012)Gerbode, Puzey, McCormick, and
  Mahadevan]{gerbode2012cucumber}
Gerbode,~S.~J. \latin{et~al.}  How the cucumber tendril coils and overwinds.
  \emph{Science} \textbf{2012}, \emph{337}, 1087--1091\relax
\mciteBstWouldAddEndPuncttrue
\mciteSetBstMidEndSepPunct{\mcitedefaultmidpunct}
{\mcitedefaultendpunct}{\mcitedefaultseppunct}\relax
\EndOfBibitem
\bibitem[Lauga and Powers(2009)Lauga, and Powers]{lauga2009hydrodynamics}
Lauga,~E. \latin{et~al.}  The hydrodynamics of swimming microorganisms.
  \emph{Reports on Progress in Physics} \textbf{2009}, \emph{72}, 096601\relax
\mciteBstWouldAddEndPuncttrue
\mciteSetBstMidEndSepPunct{\mcitedefaultmidpunct}
{\mcitedefaultendpunct}{\mcitedefaultseppunct}\relax
\EndOfBibitem
\bibitem[Purcell(1977)]{purcell1977life}
Purcell,~E.~M. Life at low Reynolds number. \emph{American journal of physics}
  \textbf{1977}, \emph{45}, 3--11\relax
\mciteBstWouldAddEndPuncttrue
\mciteSetBstMidEndSepPunct{\mcitedefaultmidpunct}
{\mcitedefaultendpunct}{\mcitedefaultseppunct}\relax
\EndOfBibitem
\bibitem[Liu \latin{et~al.}(2014)Liu, Zhang, Kim, and Park]{liu2014helical}
Liu,~L. \latin{et~al.}  Helical metallic micro-and nanostructures: fabrication
  and application. \emph{Nanoscale} \textbf{2014}, \emph{6}, 9355--9365\relax
\mciteBstWouldAddEndPuncttrue
\mciteSetBstMidEndSepPunct{\mcitedefaultmidpunct}
{\mcitedefaultendpunct}{\mcitedefaultseppunct}\relax
\EndOfBibitem
\bibitem[Nelson \latin{et~al.}(2010)Nelson, Kaliakatsos, and
  Abbott]{nelson2010microrobots}
Nelson,~B.~J. \latin{et~al.}  Microrobots for minimally invasive medicine.
  \emph{Annual review of biomedical engineering} \textbf{2010}, \emph{12},
  55--85\relax
\mciteBstWouldAddEndPuncttrue
\mciteSetBstMidEndSepPunct{\mcitedefaultmidpunct}
{\mcitedefaultendpunct}{\mcitedefaultseppunct}\relax
\EndOfBibitem
\bibitem[Attia \latin{et~al.}(2009)Attia, Pregibon, Doyle, Viovy, and
  Bartolo]{attia2009soft}
Attia,~R. \latin{et~al.}  Soft microflow sensors. \emph{Lab on a Chip}
  \textbf{2009}, \emph{9}, 1213--1218\relax
\mciteBstWouldAddEndPuncttrue
\mciteSetBstMidEndSepPunct{\mcitedefaultmidpunct}
{\mcitedefaultendpunct}{\mcitedefaultseppunct}\relax
\EndOfBibitem
\bibitem[Li \latin{et~al.}(2012)Li, Huang, Wang, Yu, Wu, Cui, and
  Mei]{li2012superelastic}
Li,~W. \latin{et~al.}  Superelastic metal microsprings as fluidic sensors and
  actuators. \emph{Lab on a Chip} \textbf{2012}, \emph{12}, 2322--2328\relax
\mciteBstWouldAddEndPuncttrue
\mciteSetBstMidEndSepPunct{\mcitedefaultmidpunct}
{\mcitedefaultendpunct}{\mcitedefaultseppunct}\relax
\EndOfBibitem
\bibitem[Tottori \latin{et~al.}(2012)Tottori, Zhang, Qiu, Krawczyk,
  Franco-Obreg{\'o}n, and Nelson]{tottori2012magnetic}
Tottori,~S. \latin{et~al.}  Magnetic helical micromachines: fabrication,
  controlled swimming, and cargo transport. \emph{Advanced materials}
  \textbf{2012}, \emph{24}, 811--816\relax
\mciteBstWouldAddEndPuncttrue
\mciteSetBstMidEndSepPunct{\mcitedefaultmidpunct}
{\mcitedefaultendpunct}{\mcitedefaultseppunct}\relax
\EndOfBibitem
\bibitem[Lu \latin{et~al.}(2018)Lu, Wang, Tan, Yang, and
  Shen]{lu2018nanorobotic}
Lu,~H. \latin{et~al.}  Nanorobotic system for precise in situ three-dimensional
  manufacture of helical microstructures. \emph{IEEE Robotics and Automation
  Letters} \textbf{2018}, \emph{3}, 2846--2853\relax
\mciteBstWouldAddEndPuncttrue
\mciteSetBstMidEndSepPunct{\mcitedefaultmidpunct}
{\mcitedefaultendpunct}{\mcitedefaultseppunct}\relax
\EndOfBibitem
\bibitem[Silva \latin{et~al.}(2017)Silva, de~Abreu, and
  Godinho]{silva2017shaping}
Silva,~P. \latin{et~al.}  Shaping helical electrospun filaments: a review.
  \emph{Soft Matter} \textbf{2017}, \emph{13}, 6678--6688\relax
\mciteBstWouldAddEndPuncttrue
\mciteSetBstMidEndSepPunct{\mcitedefaultmidpunct}
{\mcitedefaultendpunct}{\mcitedefaultseppunct}\relax
\EndOfBibitem
\bibitem[Zhang \latin{et~al.}(2009)Zhang, Abbott, Dong, Kratochvil, Bell, and
  Nelson]{zhang2009artificial}
Zhang,~L. \latin{et~al.}  Artificial bacterial flagella: Fabrication and
  magnetic control. \emph{Applied Physics Letters} \textbf{2009}, \emph{94},
  064107\relax
\mciteBstWouldAddEndPuncttrue
\mciteSetBstMidEndSepPunct{\mcitedefaultmidpunct}
{\mcitedefaultendpunct}{\mcitedefaultseppunct}\relax
\EndOfBibitem
\bibitem[Huang \latin{et~al.}(2005)Huang, Boone, Roberts, Savage, Lagally,
  Shaji, Qin, Blick, Nairn, and Liu]{huang2005nanomechanical}
Huang,~M. \latin{et~al.}  Nanomechanical architecture of strained bilayer thin
  films: from design principles to experimental fabrication. \emph{Advanced
  Materials} \textbf{2005}, \emph{17}, 2860--2864\relax
\mciteBstWouldAddEndPuncttrue
\mciteSetBstMidEndSepPunct{\mcitedefaultmidpunct}
{\mcitedefaultendpunct}{\mcitedefaultseppunct}\relax
\EndOfBibitem
\bibitem[Zhang \latin{et~al.}(2017)Zhang, Mourran, and
  Moller]{zhang2017dynamic}
Zhang,~H. \latin{et~al.}  Dynamic switching of helical microgel ribbons.
  \emph{Nano letters} \textbf{2017}, \emph{17}, 2010--2014\relax
\mciteBstWouldAddEndPuncttrue
\mciteSetBstMidEndSepPunct{\mcitedefaultmidpunct}
{\mcitedefaultendpunct}{\mcitedefaultseppunct}\relax
\EndOfBibitem
\bibitem[Jeong \latin{et~al.}(2017)Jeong, Cho, Lee, Gong, Kamien, Yang, and
  Yodh]{jeong2017topography}
Jeong,~J. \latin{et~al.}  Topography-guided buckling of swollen polymer bilayer
  films into three-dimensional structures. \emph{Soft matter} \textbf{2017},
  \emph{13}, 956--962\relax
\mciteBstWouldAddEndPuncttrue
\mciteSetBstMidEndSepPunct{\mcitedefaultmidpunct}
{\mcitedefaultendpunct}{\mcitedefaultseppunct}\relax
\EndOfBibitem
\bibitem[Liu \latin{et~al.}(2014)Liu, Huang, Su, Bertoldi, and
  Clarke]{liu2014structural}
Liu,~J. \latin{et~al.}  Structural transition from helices to hemihelices.
  \emph{PloS one} \textbf{2014}, \emph{9}\relax
\mciteBstWouldAddEndPuncttrue
\mciteSetBstMidEndSepPunct{\mcitedefaultmidpunct}
{\mcitedefaultendpunct}{\mcitedefaultseppunct}\relax
\EndOfBibitem
\bibitem[Prior \latin{et~al.}(2016)Prior, Moussou, Chakrabarti, Jensen, and
  Juel]{prior2016ribbon}
Prior,~C. \latin{et~al.}  Ribbon curling via stress relaxation in thin polymer
  films. \emph{Proceedings of the National Academy of Sciences} \textbf{2016},
  \emph{113}, 1719--1724\relax
\mciteBstWouldAddEndPuncttrue
\mciteSetBstMidEndSepPunct{\mcitedefaultmidpunct}
{\mcitedefaultendpunct}{\mcitedefaultseppunct}\relax
\EndOfBibitem
\bibitem[Douezan \latin{et~al.}(2011)Douezan, Wyart, Brochard-Wyart, and
  Cuvelier]{douezan2011curling}
Douezan,~S. \latin{et~al.}  Curling instability induced by swelling. \emph{Soft
  Matter} \textbf{2011}, \emph{7}, 1506--1511\relax
\mciteBstWouldAddEndPuncttrue
\mciteSetBstMidEndSepPunct{\mcitedefaultmidpunct}
{\mcitedefaultendpunct}{\mcitedefaultseppunct}\relax
\EndOfBibitem
\bibitem[Yu \latin{et~al.}(2017)Yu, Zhang, Hu, Grover, Huang, Wang, and
  Chen]{yu2017shape}
Yu,~X. \latin{et~al.}  Shape formation of helical ribbons induced by material
  anisotropy. \emph{Applied Physics Letters} \textbf{2017}, \emph{110},
  091901\relax
\mciteBstWouldAddEndPuncttrue
\mciteSetBstMidEndSepPunct{\mcitedefaultmidpunct}
{\mcitedefaultendpunct}{\mcitedefaultseppunct}\relax
\EndOfBibitem
\bibitem[Pham \latin{et~al.}(2013)Pham, Lawrence, Lee, Grason, Emrick, and
  Crosby]{pham2013highly}
Pham,~J.~T. \latin{et~al.}  Highly stretchable nanoparticle helices through
  geometric asymmetry and surface forces. \emph{Advanced Materials}
  \textbf{2013}, \emph{25}, 6703--6708\relax
\mciteBstWouldAddEndPuncttrue
\mciteSetBstMidEndSepPunct{\mcitedefaultmidpunct}
{\mcitedefaultendpunct}{\mcitedefaultseppunct}\relax
\EndOfBibitem
\bibitem[Maier \latin{et~al.}(2017)Maier, Bae, Schiffels, Emmerig, Schiff, and
  Liedl]{maier2017self}
Maier,~A.~M. \latin{et~al.}  Self-assembled DNA tubes forming helices of
  controlled diameter and chirality. \emph{ACS nano} \textbf{2017}, \emph{11},
  1301--1306\relax
\mciteBstWouldAddEndPuncttrue
\mciteSetBstMidEndSepPunct{\mcitedefaultmidpunct}
{\mcitedefaultendpunct}{\mcitedefaultseppunct}\relax
\EndOfBibitem
\bibitem[Kim \latin{et~al.}(2010)Kim, Lee, Sudeep, Emrick, and
  Crosby]{kim2010nanoparticle}
Kim,~H.~S. \latin{et~al.}  Nanoparticle stripes, grids, and ribbons produced by
  flow coating. \emph{Advanced Materials} \textbf{2010}, \emph{22},
  4600--4604\relax
\mciteBstWouldAddEndPuncttrue
\mciteSetBstMidEndSepPunct{\mcitedefaultmidpunct}
{\mcitedefaultendpunct}{\mcitedefaultseppunct}\relax
\EndOfBibitem
\bibitem[Choudhary and Crosby(2019)Choudhary, and
  Crosby]{choudhary2019controlled}
Choudhary,~S. \latin{et~al.}  Controlled processing of polymer nanoribbons:
  Enabling microhelix transformations. \emph{Journal of Polymer Science Part B:
  Polymer Physics} \textbf{2019}, \emph{57}, 1270--1278\relax
\mciteBstWouldAddEndPuncttrue
\mciteSetBstMidEndSepPunct{\mcitedefaultmidpunct}
{\mcitedefaultendpunct}{\mcitedefaultseppunct}\relax
\EndOfBibitem
\bibitem[Lee \latin{et~al.}(2013)Lee, Pham, Lawrence, Lee, Parkos, Emrick, and
  Crosby]{lee2013macroscopic}
Lee,~D.~Y. \latin{et~al.}  Macroscopic nanoparticle ribbons and fabrics.
  \emph{Advanced materials} \textbf{2013}, \emph{25}, 1248--1253\relax
\mciteBstWouldAddEndPuncttrue
\mciteSetBstMidEndSepPunct{\mcitedefaultmidpunct}
{\mcitedefaultendpunct}{\mcitedefaultseppunct}\relax
\EndOfBibitem
\bibitem[Gloumakoff and Yu(1964)Gloumakoff, and Yu]{gloumakoff1964}
Gloumakoff,~N.~A. \latin{et~al.}  {Torsion of Bars With Isosceles Triangular
  and Diamond Sections}. \emph{Journal of Applied Mechanics} \textbf{1964},
  \emph{31}, 332--334\relax
\mciteBstWouldAddEndPuncttrue
\mciteSetBstMidEndSepPunct{\mcitedefaultmidpunct}
{\mcitedefaultendpunct}{\mcitedefaultseppunct}\relax
\EndOfBibitem
\bibitem[Mahadevan and Keller(1993)Mahadevan, and Keller]{mahadevan1993shape}
Mahadevan,~L. \latin{et~al.}  The shape of a M{\"o}bius band. \emph{Proceedings
  of the Royal Society of London. Series A: Mathematical and Physical Sciences}
  \textbf{1993}, \emph{440}, 149--162\relax
\mciteBstWouldAddEndPuncttrue
\mciteSetBstMidEndSepPunct{\mcitedefaultmidpunct}
{\mcitedefaultendpunct}{\mcitedefaultseppunct}\relax
\EndOfBibitem
\bibitem[Starostin and Van Der~Heijden(2007)Starostin, and Van
  Der~Heijden]{starostin2007shape}
Starostin,~E. \latin{et~al.}  The shape of a M{\"o}bius strip. \emph{Nature
  materials} \textbf{2007}, \emph{6}, 563--567\relax
\mciteBstWouldAddEndPuncttrue
\mciteSetBstMidEndSepPunct{\mcitedefaultmidpunct}
{\mcitedefaultendpunct}{\mcitedefaultseppunct}\relax
\EndOfBibitem
\bibitem[Love(1944)]{love2013treatise}
Love,~A. E.~H. \emph{A treatise on the mathematical theory of elasticity};
  Cambridge university press, 1944; Chapter 19, pp 413--416\relax
\mciteBstWouldAddEndPuncttrue
\mciteSetBstMidEndSepPunct{\mcitedefaultmidpunct}
{\mcitedefaultendpunct}{\mcitedefaultseppunct}\relax
\EndOfBibitem
\bibitem[Starostin and van~der Heijden(2008)Starostin, and van~der
  Heijden]{starostin2008tension}
Starostin,~E. \latin{et~al.}  Tension-induced multistability in inextensible
  helical ribbons. \emph{Physical review letters} \textbf{2008}, \emph{101},
  084301\relax
\mciteBstWouldAddEndPuncttrue
\mciteSetBstMidEndSepPunct{\mcitedefaultmidpunct}
{\mcitedefaultendpunct}{\mcitedefaultseppunct}\relax
\EndOfBibitem
\bibitem[Pham \latin{et~al.}(2015)Pham, Morozov, Crosby, Lindner, and
  du~Roure]{pham2015deformation}
Pham,~J.~T. \latin{et~al.}  Deformation and shape of flexible, microscale
  helices in viscous flow. \emph{Physical Review E} \textbf{2015}, \emph{92},
  011004\relax
\mciteBstWouldAddEndPuncttrue
\mciteSetBstMidEndSepPunct{\mcitedefaultmidpunct}
{\mcitedefaultendpunct}{\mcitedefaultseppunct}\relax
\EndOfBibitem
\bibitem[Darnton and Berg(2007)Darnton, and Berg]{darnton2007force}
Darnton,~N.~C. \latin{et~al.}  Force-extension measurements on bacterial
  flagella: triggering polymorphic transformations. \emph{Biophysical journal}
  \textbf{2007}, \emph{92}, 2230--2236\relax
\mciteBstWouldAddEndPuncttrue
\mciteSetBstMidEndSepPunct{\mcitedefaultmidpunct}
{\mcitedefaultendpunct}{\mcitedefaultseppunct}\relax
\EndOfBibitem
\bibitem[Li \latin{et~al.}(2021)Li, Gompper, and Ripoll]{li2021tumbling}
Li,~R. \latin{et~al.}  Tumbling and Vorticity Drift of Flexible Helicoidal
  Polymers in Shear Flow. \emph{Macromolecules} \textbf{2021}, \emph{54},
  812--823\relax
\mciteBstWouldAddEndPuncttrue
\mciteSetBstMidEndSepPunct{\mcitedefaultmidpunct}
{\mcitedefaultendpunct}{\mcitedefaultseppunct}\relax
\EndOfBibitem
\bibitem[Rodenborn \latin{et~al.}(2013)Rodenborn, Chen, Swinney, Liu, and
  Zhang]{rodenborn2013propulsion}
Rodenborn,~B. \latin{et~al.}  Propulsion of microorganisms by a helical
  flagellum. \emph{Proceedings of the National Academy of Sciences}
  \textbf{2013}, \emph{110}, E338--E347\relax
\mciteBstWouldAddEndPuncttrue
\mciteSetBstMidEndSepPunct{\mcitedefaultmidpunct}
{\mcitedefaultendpunct}{\mcitedefaultseppunct}\relax
\EndOfBibitem
\bibitem[Ferry(1980)]{ferry1980viscoelastic}
Ferry,~J.~D. \emph{Viscoelastic properties of polymers}; John Wiley \& Sons,
  1980\relax
\mciteBstWouldAddEndPuncttrue
\mciteSetBstMidEndSepPunct{\mcitedefaultmidpunct}
{\mcitedefaultendpunct}{\mcitedefaultseppunct}\relax
\EndOfBibitem
\bibitem[Roth and Dutcher(2005)Roth, and Dutcher]{roth2005glass}
Roth,~C.~B. \latin{et~al.}  Glass transition and chain mobility in thin polymer
  films. \emph{Journal of Electroanalytical Chemistry} \textbf{2005},
  \emph{584}, 13--22\relax
\mciteBstWouldAddEndPuncttrue
\mciteSetBstMidEndSepPunct{\mcitedefaultmidpunct}
{\mcitedefaultendpunct}{\mcitedefaultseppunct}\relax
\EndOfBibitem
\bibitem[Pye and Roth(2011)Pye, and Roth]{pye2011two}
Pye,~J.~E. \latin{et~al.}  Two simultaneous mechanisms causing glass transition
  temperature reductions in high molecular weight freestanding polymer films as
  measured by transmission ellipsometry. \emph{Physical review letters}
  \textbf{2011}, \emph{107}, 235701\relax
\mciteBstWouldAddEndPuncttrue
\mciteSetBstMidEndSepPunct{\mcitedefaultmidpunct}
{\mcitedefaultendpunct}{\mcitedefaultseppunct}\relax
\EndOfBibitem
\bibitem[O'Connell \latin{et~al.}(2008)O'Connell, Hutcheson, and
  McKenna]{o2008creep}
O'Connell,~P.~A. \latin{et~al.}  Creep behavior of ultra-thin polymer films.
  \emph{Journal of Polymer Science Part B: Polymer Physics} \textbf{2008},
  \emph{46}, 1952--1965\relax
\mciteBstWouldAddEndPuncttrue
\mciteSetBstMidEndSepPunct{\mcitedefaultmidpunct}
{\mcitedefaultendpunct}{\mcitedefaultseppunct}\relax
\EndOfBibitem
\bibitem[Mattsson \latin{et~al.}(2000)Mattsson, Forrest, and
  B{\"o}rjesson]{mattsson2000quantifying}
Mattsson,~J. \latin{et~al.}  Quantifying glass transition behavior in ultrathin
  free-standing polymer films. \emph{Physical Review E} \textbf{2000},
  \emph{62}, 5187\relax
\mciteBstWouldAddEndPuncttrue
\mciteSetBstMidEndSepPunct{\mcitedefaultmidpunct}
{\mcitedefaultendpunct}{\mcitedefaultseppunct}\relax
\EndOfBibitem
\bibitem[Rowland \latin{et~al.}(2008)Rowland, King, Cross, and
  Pethica]{rowland2008measuring}
Rowland,~H.~D. \latin{et~al.}  Measuring glassy and viscoelastic polymer flow
  in molecular-scale gaps using a flat punch mechanical probe. \emph{ACS nano}
  \textbf{2008}, \emph{2}, 419--428\relax
\mciteBstWouldAddEndPuncttrue
\mciteSetBstMidEndSepPunct{\mcitedefaultmidpunct}
{\mcitedefaultendpunct}{\mcitedefaultseppunct}\relax
\EndOfBibitem
\bibitem[Bodiguel and Fretigny(2006)Bodiguel, and
  Fretigny]{bodiguel2006reduced}
Bodiguel,~H. \latin{et~al.}  Reduced viscosity in thin polymer films.
  \emph{Physical review letters} \textbf{2006}, \emph{97}, 266105\relax
\mciteBstWouldAddEndPuncttrue
\mciteSetBstMidEndSepPunct{\mcitedefaultmidpunct}
{\mcitedefaultendpunct}{\mcitedefaultseppunct}\relax
\EndOfBibitem
\bibitem[Li \latin{et~al.}(2013)Li, Chen, Lam, and Tsui]{li2013viscosity}
Li,~R.~N. \latin{et~al.}  Viscosity of PMMA on silica: Epitome of systems with
  strong polymer--substrate interactions. \emph{Macromolecules} \textbf{2013},
  \emph{46}, 7889--7893\relax
\mciteBstWouldAddEndPuncttrue
\mciteSetBstMidEndSepPunct{\mcitedefaultmidpunct}
{\mcitedefaultendpunct}{\mcitedefaultseppunct}\relax
\EndOfBibitem
\bibitem[Sussman \latin{et~al.}(2014)Sussman, Tung, Winey, Schweizer, and
  Riggleman]{sussman2014entanglement}
Sussman,~D.~M. \latin{et~al.}  Entanglement reduction and anisotropic chain and
  primitive path conformations in polymer melts under thin film and cylindrical
  confinement. \emph{Macromolecules} \textbf{2014}, \emph{47}, 6462--6472\relax
\mciteBstWouldAddEndPuncttrue
\mciteSetBstMidEndSepPunct{\mcitedefaultmidpunct}
{\mcitedefaultendpunct}{\mcitedefaultseppunct}\relax
\EndOfBibitem
\bibitem[Si \latin{et~al.}(2005)Si, Massa, Dalnoki-Veress, Brown, and
  Jones]{si2005chain}
Si,~L. \latin{et~al.}  Chain entanglement in thin freestanding polymer films.
  \emph{Physical review letters} \textbf{2005}, \emph{94}, 127801\relax
\mciteBstWouldAddEndPuncttrue
\mciteSetBstMidEndSepPunct{\mcitedefaultmidpunct}
{\mcitedefaultendpunct}{\mcitedefaultseppunct}\relax
\EndOfBibitem
\bibitem[Martin \latin{et~al.}(2010)Martin, Krutyeva, Monkenbusch, Arbe,
  Allgaier, Radulescu, Falus, Maiz, Mijangos, Colmenero, \latin{et~al.}
  others]{martin2010direct}
Martin,~J. \latin{et~al.}  Direct observation of confined single chain dynamics
  by neutron scattering. \emph{Physical review letters} \textbf{2010},
  \emph{104}, 197801\relax
\mciteBstWouldAddEndPuncttrue
\mciteSetBstMidEndSepPunct{\mcitedefaultmidpunct}
{\mcitedefaultendpunct}{\mcitedefaultseppunct}\relax
\EndOfBibitem
\bibitem[Barber \latin{et~al.}(2020)Barber, Yang, Pr{\'e}vost, Du~Roure,
  Lindner, Emrick, and Crosby]{barber2020programmed}
Barber,~D.~M. \latin{et~al.}  Programmed Wrapping and Assembly of Droplets with
  Mesoscale Polymers. \emph{Advanced Functional Materials} \textbf{2020},
  \emph{30}, 2002704\relax
\mciteBstWouldAddEndPuncttrue
\mciteSetBstMidEndSepPunct{\mcitedefaultmidpunct}
{\mcitedefaultendpunct}{\mcitedefaultseppunct}\relax
\EndOfBibitem
\end{mcitethebibliography}

\end{document}